\documentclass[12pt]{iopart}

\usepackage{amssymb}
\usepackage[figuresright]{rotating}
\begin{document}

\title[]{Hybrid model of GeV-TeV gamma ray emission from Galactic Center}

\author{Yi-Qing Guo$^1$, Qiang Yuan$^{1,2}$, Cheng Liu$^1$, Ai-Feng Li$^3$}

\address{$^1$Key Laboratory of Particle Astrophysics, Institute of High 
Energy Physics, Chinese Academy of Science, Beijing 100049, P. R. China\\
$^2$Key Laboratory of Dark Matter and Space Astronomy, Purple Mountain
Observatory, Chinese Academy of Sciences, Nanjing 210008, P. R. China\\
$^3$ School of Information Science and Engineering, Shandong 
Agriculture University, Taian ,271018, China
}

\ead{yuanq@ihep.ac.cn}

\begin{abstract}
The observations of high energy $\gamma$-ray emission from the Galactic 
center (GC) by HESS, and recently by Fermi, suggest the cosmic ray 
acceleration in the GC and possibly around the supermassive black hole.
In this work we propose a lepton-hadron hybrid model to explain 
simultaneously the GeV-TeV $\gamma$-ray emission. Both electrons and
hadronic cosmic rays were accelerated during the past activity of the
GC. Then these particles would diffuse outwards and interact with the
interstellar gas and background radiation field. The collisions between 
hadronic cosmic rays with gas is responsible to the TeV $\gamma$-ray 
emission detected by HESS. With fast cooling in the strong radiation
field, the electrons would cool down and radiate GeV photons through
inverse Compton scattering off the soft background photons. This 
scenario provides a natural explanation of the observed GeV-TeV
spectral shape of $\gamma$-rays. 
\end{abstract}
\maketitle

\section{Introduction}

It is well known that the Galactic center (GC) region has a very complex
astrophysical enviroment and is rich in various kinds of objects, and is 
a good library for the study of astrophysical phonomena. A supermassive 
black hole with mass $\sim 4\times10^6$ M$_{\odot}$, Sgr A$^{\star}$, 
lies in the GC. Although the GC is rather quiet nowadays, frequent flares 
were observed in the X-ray as well as near infrared (NIR) bands 
\cite{2009ApJ...698..676D,2011ApJ...728...37D,2012ApJ...759...95N}, 
which means the existence of continuous weak activities of the black hole. 

As a consequence high energy particles could be accelerated during 
such kinds of activities and imprinted in the $\gamma$-ray sky. The 
very high energy (VHE) $\gamma$-rays from the GC have been observed 
by several atmospheric Cerenkov telescopes such as CANGAROO 
\cite{2004ApJ...606L.115T}, VERITAS \cite{2004ApJ...608L..97K}, 
HESS \cite{2004A&A...425L..13A,2006PhRvL..97x9901A,2008A&A...492L..25A}, 
and MAGIC \cite{2006ApJ...638L.101A}. Recently GeV $\gamma$-ray 
emission from the GC was also revealed by the spatial detector Fermi 
Large Area Telescope (Fermi-LAT, \cite{2011ApJ...726...60C}). 
The observations in both TeV and GeV bands showed that the $\gamma$-ray 
emission tended to be stable without significant variability 
\cite{2008A&A...492L..25A,2006ApJ...638L.101A,2011ApJ...726...60C}. 

Serveral models have been proposed to explain the $\gamma$-ray emission 
observed at the GC, including hadronic models 
\cite{2004ApJ...617L.123A,2006ApJ...647.1099L,2007ApJ...657L..13B,
2011ApJ...726...60C,2012ApJ...753...41L,2012ApJ...757L..16F} 
and leptonic ones \cite{2007ApJ...657..302H,2012ApJ...748...34K}.
An issue prevents a simple explanation to the GeV-TeV spectra of
the GC is that the GeV $\gamma$-ray emission is not consistent with
the direct extrapolation of TeV $\gamma$-rays to the low energy range.
Chernyakova et al. \cite{2011ApJ...726...60C} proposed that the strong 
energy dependence of the diffusion coefficient would seperate the low 
energy and high energy particles into two different diffusion regimes 
and the GeV-TeV spectral shape of the $\gamma$-rays could be reproduced. 
In \cite{2012ApJ...757L..16F} the stochastic acceleration of particles
in a two-phase interstellar medium (ISM) was employed to generate two 
distinct populations of protons to explain the data. 
In \cite{2012ApJ...748...34K} the Fermi-LAT $\gamma$-ray emission was 
suggested to originate from a population of electrons ICS off the soft 
background photons and the VHE $\gamma$-ray emission was thought to be 
from different sources instead of Sgr A$^{\star}$.

The morphology study of the VHE $\gamma$-ray source, HESS J1745-290, by 
HESS showed no spatial extension and the upper limit of the angular size
of $\sim 1'.3$ was set \cite{2010MNRAS.402.1877A}. Note there is another
diffuse component of the $\gamma$-ray emission along the ridge
\cite{2006Natur.439..695A,2012arXiv1206.6882Y}, which is not discussed 
in this work. It may imply that the high energy particles are confined 
in several parsec regions around the GC \cite{2011ApJ...726...60C,
2012ApJ...757L..16F} or the high concentration of the target gas in the 
inner region \cite{2012ApJ...753...41L}.

In this work, we propose a hybrid model to explain the GeV and TeV 
$\gamma$-ray emission. It is natural to expect that both protons and 
electrons can be accelerated during the GC activity. Therefore without
finely tuning the environmental parameters, we should expect the 
existence of two components of the $\gamma$-ray emission, which is
just the case shown by the GeV-TeV observations. A simple expectation
is that high energy protons interact with the ISM is responsible to 
the VHE $\gamma$-rays, and the electrons, which may cool down in the 
GC radiation field and/or magnetic field, produce the GeV $\gamma$-rays 
through inverse Compton scattering (ICS) off the background photons. 
In Sec. 2, we present the detailed modeling of this picture. 
Sec. 3 is the discussion and the main conclusions are summarized in Sec. 4.

\section{Model and results}
\label{sect:Model}

Although the overall behavior of Sgr A$^{\star}$ is quite silent, the 
observations in X-ray and infrared bands indicate that it still has
continuous weak activities \cite{1982ApJ...258..135B,1992ApJ...387..189D,
2009ApJ...698..676D,2011ApJ...728...37D}. During such kinds of avtivities, 
the accretion of stars and gas by the supermassive black hole could be 
effictive to accelerate particles. We assume that both the protons and 
electrons were accelerated instantaneously during such flare events. Then 
these particles would diffuse away from the acceleration site, and radiate
during the propagation. The propagation equation for both protons and
electrons can be written as
\begin{equation}
\frac{\partial \phi}{\partial t}=\frac{D(E)}{r^2}\frac{\partial}
{\partial r}r^2\frac{\partial \phi}{\partial r}+\frac{\partial}
{\partial E}\left[b(E)\phi\right]+N(E)\delta(t)\delta({\bf r}),
\end{equation}
where $\phi(r,E,t)$ is the propagated flux as a function of space, energy
and time, $D(E)$ is the diffusion coefficient, $N(E)$ is the injection
spectrum of particles at $t=0$ and ${\bf r}=0$, and $b(E)\equiv
{\rm d}E/{\rm d}t$ is the energy loss rate, which is important for 
electrons but negligible for protons. The energy dependence of diffusion
coefficient is assumed to be power-law of rigidity $R$, $D(R(E))=\beta 
D_0(R/4\,{\rm GV})^{\delta}$, in which $\delta$ is the spectral index
and $\beta$ is the particle velocity. In this work we assume 
$\delta=0.3$, which is close to a Kolmogorov type of the ISM turbulence.

The solution of the above equation for electrons is \cite{1995PhRvD..52.3265A}
\begin{equation}
\phi_e(r,E,t)=\frac{N_e(E_i)b(E_i)}{\pi^{3/2}b(E)r_{\rm dif}^3}\exp\left(
-\frac{r^2}{r_{\rm dif}^2}\right),
\end{equation}
where $E_i$ is the initial energy of the electron whose energy is cooled
down to $E$ at time $t$, and $r_{\rm dif}(E,t)=2\sqrt{\Delta u}$, in which 
\begin{equation}
\Delta u(E,E_i)=\int_E^{E_i}\frac{D(E')}{b(E')}{\rm d}E',
\end{equation}
is the effective diffusion length when the particle energy decreases 
from $E_i$ to $E$.

For protons we neglect the energy loss term, and the solution of the
propagation equation is even simpler
\begin{equation}
\phi_p(r,E,t)=\frac{N_p(E)}{\left(\sqrt{2\pi}\sigma\right)^3}\exp
\left(-\frac{r^2}{2\sigma^2}\right),
\end{equation}
where $\sigma(E,t)=\sqrt{2D(E)t}$ is the effective diffusion length 
within $t$.

\begin{figure}[!htb]
\centering
\includegraphics[width=0.8\textwidth,angle=-0]{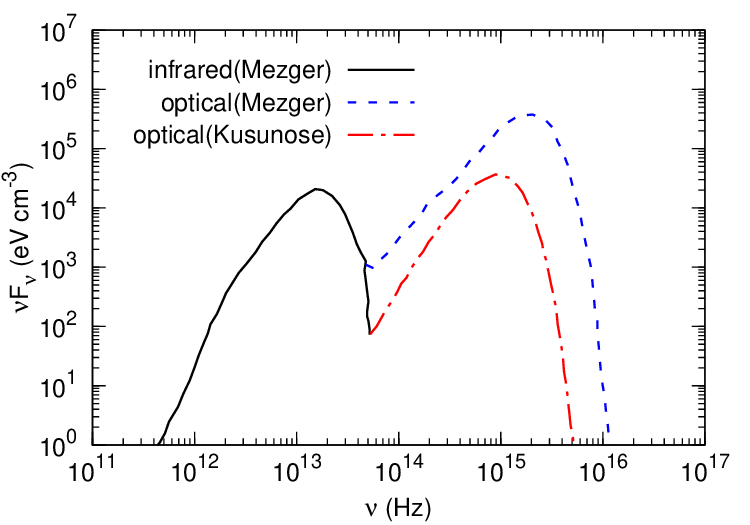}
\caption{Infrared and optical photon energy density spectrum in $\sim1$ pc
of the GC region. The solid line is adopted from \cite{1996A&ARv...7..289M} 
which is consistent with the observational data \cite{1992ApJ...387..189D}.
The dashed and dash-dotted lines are the model expected results of optical
emission presented in \cite{1996A&ARv...7..289M} and 
\cite{2012ApJ...748...34K} respectively. This figure is a reproduction of
Fig. 1 in \cite{2012ApJ...748...34K}.}
\label{bkg_Photon_Filed}
\end{figure}

The energy loss of electrons includes ICS off the soft background photons,
synchrotron radiation in the magnetic field and bremsstrahlung radiation
in the ISM. The soft photons consist with infrared from dust and optical 
from stars. In \cite{1992ApJ...387..189D} the measurement of infrared 
emission in the central $1.2$ pc was presented. According to the data, 
the expected optical emission was proposed based on assumptions of the 
optical photon energy density \cite{1996A&ARv...7..289M,
2007ApJ...657..302H,2012ApJ...748...34K}. Taking the uncertainties of the 
expected optical emission into account, we adopt the two models given in
\cite{1996A&ARv...7..289M,2012ApJ...748...34K} in this work. The intensity
of the soft photons is shown in Figure \ref{bkg_Photon_Filed}. 
The magnetic field strength varies with the radius away from the GC 
\cite{2009JCAP...03..009B}. For the very small region ($r\sim$ pc) the 
average magnetic field strength is adopted to be $\sim10^2\mu$G 
\cite{2009JCAP...03..009B,2012ApJ...748...34K}. As for the ISM density,
there is large uncertainty in the GC region. Typically the adopted
ISM density within $\sim$pc region is of the order $10^3$ cm$^{-3}$
\cite{2011ApJ...726...60C,2012ApJ...753...41L}. Compared with the 
ICS energy loss in the strong radiation field, the bremsstrahlung 
energy loss is negligible \cite{1998ApJ...509..212S}.

\begin{figure}[!htb]
\centering
\includegraphics[width=0.8\textwidth]{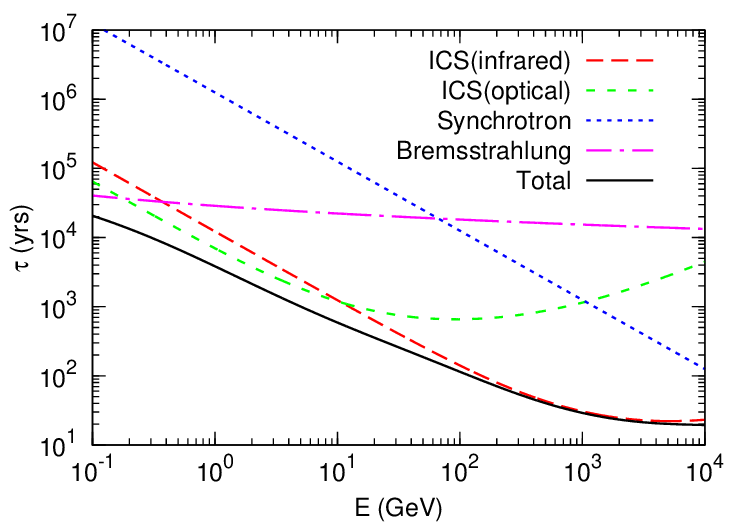}
\caption{The cooling time of electrons in the background photon field model 
of Kusunose \& Takahara (2012) \cite{2012ApJ...748...34K} and magnetic 
field with $B=100\mu$G. The ISM density is adopted to be $10^3$ cm$^{-3}$
to calculate the bremsstrahlung energy loss.}
\label{energy_loss}
\end{figure}

We calculate the cooling time of the electrons, $\tau=E/({\rm d}E/{\rm d}t)$, 
in the above background photons field and magnetic field, as shown in 
Figure \ref{energy_loss}. Here the optical emission is adopted to be the 
one of Kusunose \& Takahara (2012) \cite{2012ApJ...748...34K}. It is 
shown that the energy loss of electrons is dominated by the ICS up to 
$>10$ TeV. 

\begin{figure}[!htb]
\centering
\includegraphics[width=0.8\textwidth]{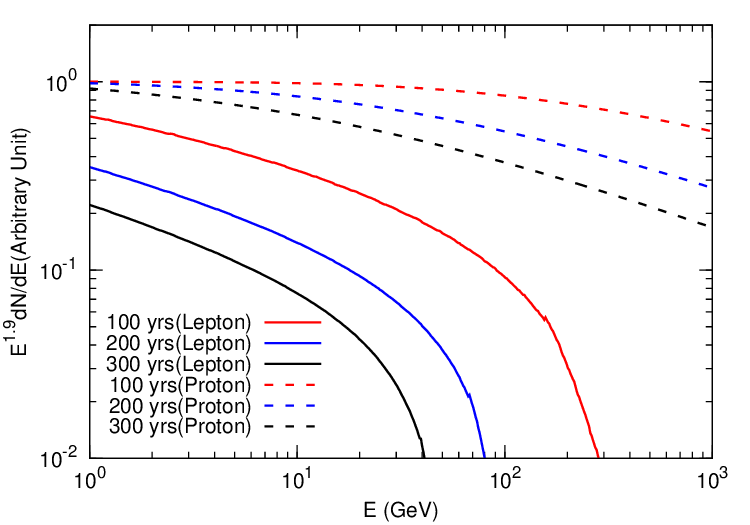}
\caption{Propagated electron (solid) and proton (dashed) spectra for 
observational time $t=100,\,200$ and $300$ (from top to bottom) years 
respectively. The energy loss of electrons adopted in this calculation
is shown in Figure \ref{energy_loss}.}
\label{spctGreenAll}
\end{figure}

In Figure \ref{spctGreenAll} we show the integral spectra of both 
electrons and protons after time $t$ of the injection. The integral 
radius is taken to be $1.2$ pc for electrons and $3$ pc for protons
respectively. For electrons we integrate the particles within radius 
$\sim 1.2$ pc since the high background photon density is expected to 
exist in that region, and most of the radiation should come from such 
a region. However, for protons, we adopt a some larger integral radius, 
$\sim 3$ pc, according to the spatial extention upper limit of 
HESS J1745-290 derived by HESS \cite{2010MNRAS.402.1877A}. The diffusion
coefficient is $D_0=10^{27}$ cm$^2$ s$^{-1}$, and the injection spectra
of both electrons and protons are adopted as $E^{-1.9}$. 
We can see from this figure that the cooling of electrons 
will result in a cutoff of the spectrum, above which the electrons 
can not survive from the energy loss. The cutoff energy depends on the 
time. The younger the injection is, the higher energy of electrons we can 
observe. The spectral shape also has an evolution with the time. For 
shorter time, the relative fraction of low energy particles is smaller
due to slower propagation of low energy particles, which results in
harder spectra. For protons there is also a time evolution of the spectra.
It reflects the fact that for longer time, more high energy particles 
will diffuse out of the integral region and result in a soft spectrum.

Then we discuss the $\gamma$-ray emission of those electrons and protons
in the background radiation field and the interstellar medium. For 
electrons we consider only the ICS emission and neglect the bremsstrahlung
radiation. The Klein-Nishina cross section of ICS is employed. For the
$\gamma$-ray emission from $pp$ collisions, we use the parameterization
given in \cite{2006ApJ...647..692K}. 

\begin{figure}[!htb]
\centering
\includegraphics[width=0.8\textwidth]{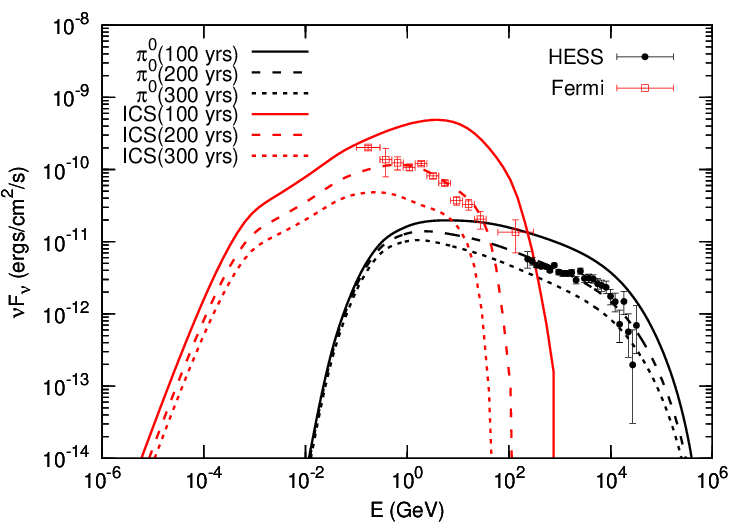}
\caption{Calculated $\gamma$-ray spectra of the hybrid model, compared
with the observational data by Fermi-LAT \cite{2011ApJ...726...60C} and 
HESS \cite{2009A&A...503..817A}. Different lines show the effect of 
observational time $t$ (100, 200 and 300 yr from top to bottom). 
The diffusion coefficient is $D_0=10^{27}$ cm$^2$ s$^{-1}$. The soft 
background photon model is Kusunose \& Takahara (2012) 
\cite{2012ApJ...748...34K}.
}
\label{spctKusunose}
\end{figure}

Figure \ref{spctKusunose} presents the results of the calculated 
$\gamma$-ray spectra and the comparison with the observational data 
\cite{2011ApJ...726...60C,2009A&A...503..817A}. The injection spectra
are adopted to be $\propto E^{-1.9}$, and the diffusion coefficient is 
adopted to be $D_0=10^{27}$ cm$^2$ s$^{-1}$. The background photon model 
is adopted to be Kusunose \& Takahara (2012) \cite{2012ApJ...748...34K}. 
Three different observational time $t=100$, $200$ and $300$ yr are
shown. The observational time determines the cutoff energy of the
electron spectrum (see Figure \ref{spctGreenAll}). We find that for
the Kusunose \& Takahara (2012) photon field $t\sim200$ yr can match
the data well\footnote{Here we do not intend to fit the data to find 
the best parameters, but just to show the rough comparison between the 
model and data. Except the first data point of Fermi-LAT, we actually 
find rather good description to the data of the model (dashed line).
There could be systematic uncertainty of the data analysis of Fermi-LAT, 
especially at low energies, and adjusting properly the model parameters
may also further improve the fit.}. Another effect of the time $t$, is 
the overall normalization. If $t$ is larger, more particles diffuse out of 
the $\sim$pc region, and we expect a decrease of the overall normalization 
of the integral particle spectra. For $t=200$ yr we find the total
energy of electrons above $1$ GeV is $\sim 5\times 10^{47}$ erg to
match the Fermi-LAT data. The electron-to-proton ratio $K_{ep}$ is
about $0.22(n/10^3\,{\rm cm}^{-3})$ in this case. Note, however,
a significant part of high energy protons ($E>10$ TeV) will diffuse 
out of the $3$ pc region in this case, as can be seen from Figure 
\ref{spctGreenAll}. Therefore to be consistent with the morphology
of the source, we may require the gas density is higher in inner $3$ 
pc region than outside, as the scenario of \cite{2012ApJ...753...41L}.

\begin{figure}[!htb]
\centering
\includegraphics[width=0.8\textwidth]{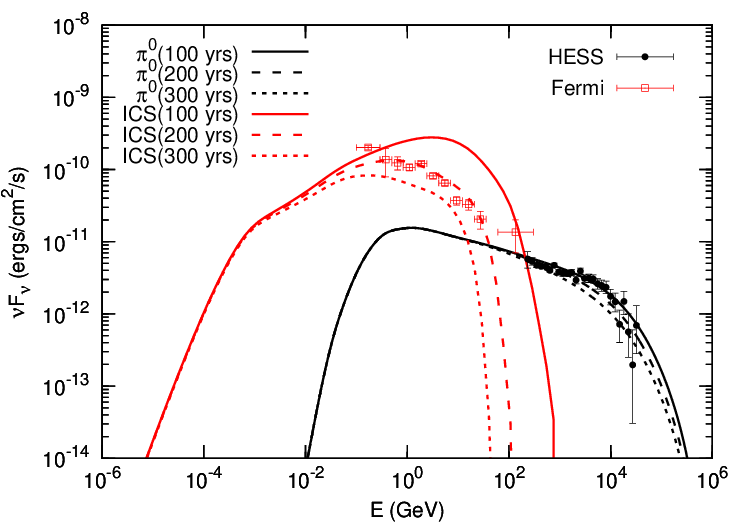}
\caption{Same as Figure \ref{spctKusunose} but for smaller diffusion 
coefficient $D_0=10^{26}$ cm$^2$ s$^{-1}$.}
\label{spctKusunoseD0}
\end{figure}

To be consistent with the cutoff behavior of VHE $\gamma$-rays observed 
by HESS, we employ an exponential cutoff of the proton spectrum with 
cutoff energy $E_c=200$ TeV. It is possible that the pair production of 
VHE photons in the background radiation field could also be responsible 
to the cutoff of the $\gamma$-ray spectra \cite{2006A&A...449..641Z,
2006ApJ...640L.155M}. However, our calculation shows that for the adopted
soft background photon field in $\sim$pc region, the pair production
attenuation is negligible. 
 Thus the cutoff should be understood as the acceleration limit of 
the flaring event. The maximum energy protons can achieve for diffusive 
shock acceleration is \cite{2005ApJ...619..306A}
\begin{equation}
E_{\rm max} \sim eBR \approx 10^{21}\left(\frac{B}{\rm G}\right)
\left(\frac{R}{\rm pc}\right)\ {\rm eV},
\end{equation}
where $B$ is the magnetic field and $R$ is the size of the acceleration
region. As in \cite{2005ApJ...619..306A}, we assume the acceleration
takes place within $10$ Schwarzschild radii ($R_g\sim10^{12}$ cm) of the 
black hole. To accelerate protons to above $100$ TeV requires a magnetic
field $\sim$G in the acceleration region. Such a condition could be
reached in the very central region of the GC \cite{2005ApJ...619..306A}.
On the other hand, if the acceleration takes place in larger regions,
the required magnetic field could be smaller.

Figure \ref{spctKusunoseD0} shows the results for even smaller diffusion
coefficient $D_0=10^{26}$ cm$^2$ s$^{-1}$. In this case most of the
protons up to $100$ TeV may still be confined in several pc region
around the GC. As a simple estimate, the diffusion length
for protons is about $1.8\left(\frac{D_0}{10^{26}\,{\rm cm^2\,s^{-1}}}
\right)^{0.5}\left(\frac{E}{\rm 100\,TeV}\right)^{0.15}\left(\frac{t}{200\,
{\rm yr}}\right)^{0.5}$ pc. The accumulative spectrum of protons is
similar with the injection spectrum. Therefore we adopt the injection
spectrum $\propto E^{-2.2}$ to be consistent with the HESS data. 
The results show a good fit to the GeV-TeV observational data for $t=200$ 
yr too. It is shown that for protons different time $t$ leads to small 
differences of the results, which means that indeed most of the protons 
are confined. We also note that the differences of the ICS spectra with
respect to $t$ are larger than that of $\pi^0$ decay spectra. This is
simply because the integral radius of electrons is smaller than that
of protons. The total energy of electrons above $1$ GeV is 
$\sim 8\times 10^{46}$ erg in this case, and the electron-to-proton ratio 
$K_{ep}$ is $0.05(n/10^3\,{\rm cm}^{-3})$. 

\begin{figure}[!htb]
\centering
\includegraphics[width=0.8\textwidth]{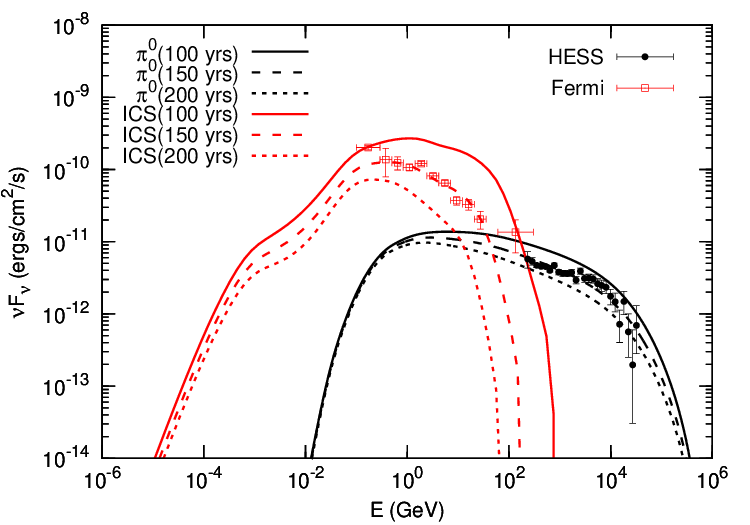}
\caption{Same as Figure \ref{spctKusunose} but for the Mezger et al. 
(1996) soft background photon model \cite{1996A&ARv...7..289M}. 
}
\label{spctMezger}
\end{figure}

Finally, the results for Mezger et al. (1996) soft background photon model 
\cite{1996A&ARv...7..289M} are shown in Figure \ref{spctMezger}. Other
conditions are the same as Figure \ref{spctKusunose}. Compared
with Kusunose \& Takahara (2012) photon field, the optical emission is
stronger and the cooling time for electrons is shorter. Thus for $t\sim
150$ yr the cutoff energy of electrons is proper to fit the Fermi-LAT
data. However, the energy spectrum of ICS emission does not differ 
significantly from that shown in Figure \ref{spctKusunose}. This is 
because in stronger background radiation field the cooling is more
significant, and a larger cooling is just canceled out when calculating
the ICS $\gamma$-ray spectrum in such a radiation field.
The total energy of electrons above $1$ GeV is $\sim 1.6\times 10^{47}$ 
erg, and the electron-to-proton ratio $K_{ep}$ is about $0.11(n/10^3\,
{\rm cm}^{-3})$.

\section{Discussion}

 The diffusion coefficient adopted in this work is $D_0\sim10^{26}-
10^{27}$ cm$^2$ s$^{-1}$, which is much smaller than that in the disk and 
halo ($\sim 10^{28}-10^{29}$ as induced from the cosmic ray transportation). 
This might be due to the higher magnetic field and more turbulent ISM
in the GC region. The lower limit of the diffusion coefficient should be 
the Bohm limit, which is $D_{\rm Bohm}=pv/3eB\sim 3\times 10^{25}
(p/{\rm TeV})(B/\mu G)^{-1}$ cm$^2$ s$^{-1}$ \cite{1981ICRC....2..336G,
1983RPPh...46..973D,1984MNRAS.207P...1H}. The diffusion coefficient 
adopted in this work is well larger than the Bohm limit in the whole 
energy range discussed.

In the calculation of proton propagation, the energy loss of protons
via interaction with the ISM is neglected. As a rough estimate, the
average collision probability of one projectile proton is about
$n\sigma v\tau\sim0.01(n/10^3\,{\rm cm}^{-3})(\tau/200\,{\rm yr})$.
Thus for the typical parameters adopted in this work, the collisional
energy loss of protons is indeed negilible. 

The secondary electrons/positrons from charged pion decay due to $pp$
collision could also contribute to the $\gamma$-ray emission. As a very 
rough estimate, for one $pp$ collision, $\gamma$-ray photons will take 
$1/3$ of the energy of all the pions, and $e^{\pm}$ will take 
$2/3\times 1/3=2/9$ of the energy (assuming energy equipartition between 
$e^{\pm}$ and neutrinos). Therefore the total energy of secondary $e^{\pm}$
should be lower than that of $\pi^0$ decaying $\gamma$-rays. 
Furthermore, the soft radiation field is restricted in 1.2 pc region
around the black hole, while the $pp$ collision occurs in $\sim 3$ pc 
region. Therefore the contribution of the secondary $e^{\pm}$ to the 
$\gamma$-rays through ICS emission should be much smaller than that of 
neutral pion decay.

The electron component will also produce synchrotron radiation in the
magnetic field, and there might be constraints from the radio to X-ray
data. As shown in \cite{2012ApJ...748...34K} the synchrotron emission
in $\sim100$ $\mu$G magnetic field may exceed the radio measurement
of Sgr A$^{\star}$ for the quiescent state \cite{2001ARA&A..39..309M}.
However, the spatial scale of the present study is arc-minutes (pc), 
instead of arc-seconds as the radio observations show. The expected
synchrotron emission is consistent with the data-based result of pc 
scale radio emission as given in \cite{1996A&ARv...7..289M}.

\begin{figure}[!htb]
\centering
\includegraphics[width=0.8\textwidth]{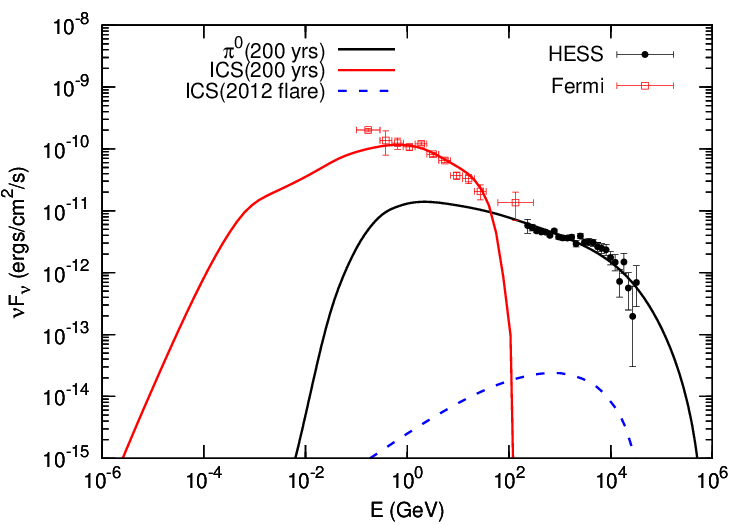}
\caption{The expected $\gamma$-ray emission for the flare event of 2012.
We assume the total power of the flare event is $4$ orders of magnitide
high than the $2-10$ keV energy.}
\label{2012Flare}
\end{figure}

Early in 2012, one bright flare of GC is observed by HETGS onboard of 
the Chandra X-ray observatory \cite{2012ApJ...759...95N}. The total 
energy in $2-10$ keV band was approximately $10^{39}$ erg
\cite{2012ApJ...759...95N}. Assuming the total energy of this accretion
event is about $4$ orders of magnitude higher \cite{1996A&ARv...7..289M}, 
$\sim10^{43}$ erg, and $\sim10\%$ of it converts to acceleration of 
cosmic rays, we give the expected $\gamma$-ray emission of such an event 
within the current framework of the model, as shown in Figure 
\ref{2012Flare}. The solid lines are same as that in Figure 
\ref{spctKusunose}. The dashed line is the expected flux of the ICS 
emission, with the same parameters of the diffusion coefficients, 
injection spectral index and electron-to-proton ratio as that of Figure
\ref{spctKusunose}. The $\gamma$-ray emission at this stage is 
dominated by the ICS emission and the hadronic component
is much lower. We can see that the total flux of $\gamma$-rays is 
too low to be able to be detected by the current VHE $\gamma$-ray 
detectors. We should keep in mind that this estimate is very rough 
and suffers from large uncertainties of the assumption of total energy 
output of the flare and the energy fraction goes into particle 
acceleration. This gives us an impression that how large a flare is 
needed in order to produce the observed $\gamma$-rays.

\section{Conclusion}

In this work, a hybrid model is proposed to explain the GeV
$\gamma$-ray emission observed by Fermi-LAT and TeV $\gamma$-ray
emission observed by HESS telescope, of the GC. The current 
$\gamma$-ray emission could be originated from one flare with 
total energy $\gtrsim10^{48}$ ergs at $\sim10^2$ years ago. 
Both the protons and electrons were accelerated to very high
energies in this flaring event. Furthermore the flaring event
produced a thermal bath of soft photons in the GC region, with
typical scale $\sim$pc. High energy particles could then diffuse out 
of the acceleration site and interact with the ambient medium and 
radiation fiels. The hadronic collisions between protons and ISM gives 
rise to the TeV $\gamma$-rays. At the same time, electrons would lose
energy through ICS and synchrotron radiation. The ICS photons 
could be responsible for the GeV $\gamma$-ray emission observed by
Fermi-LAT.

\section*{Acknowledgements}
This work is supported by the Ministry of Science and Technology of
China, Natural Sciences Foundation of China (Nos. 10725524, 10773011,
11135010 and 11105155), the Chinese Academy of Sciences (Nos. 
KJCX2-YW-N13, KJCX3-SYW-N2, GJHZ1004) and Natural Science Foundation of 
Shandong Province of China (ZR2009AM003). QY acknowledges the support 
from the Key Laboratory of Dark Matter and Space Astronomy of Chinese 
Academy of Sciences.

\end{document}